# Intraoperative perfusion assessment by continuous, low-latency hyperspectral light-field imaging: development, methodology, and clinical application

Stefan Kray*[a], Andreas Schmid[a], Eric L. Wisotzky[b,c], Moritz Gerlich[a], Sebastian Apweiler[d], Anna Hilsmann[b], Thomas Greiner[a], Peter Eisert[b,c], Werner Kneist[d],

[a]Institute of Smart Systems and Services, Pforzheim University, Pforzheim, Germany; [b]Computer Vision & Graphics, Vision & Imaging Technologies, Fraunhofer Heinrich-Hertz-Institute HHI, Berlin, Germany; [c]Visual Computing, Humboldt-University, Berlin, Germany; [d]Department of General-, Visceral- and Thoracic Surgery, Klinikum Darmstadt GmbH, Darmstadt, Germany

## ABSTRACT

Accurate assessment of tissue perfusion is crucial in visceral surgery, especially during anastomosis. Currently, subjective visual judgment is commonly employed in clinical settings. Hyperspectral imaging (HSI) offers a non-invasive, quantitative alternative. However, HSI imaging lacks continuous integration into the clinical workflow. This study presents a hyperspectral light field system for intraoperative tissue oxygen saturation ($SO_2$) analysis and visualization. We present a correlation method for determining $SO_2$ saturation with low computational demands. We demonstrate clinical application, with our results aligning with the perfusion boundaries determined by the surgeon. We perform and compare continuous perfusion analysis using two hyperspectral cameras (Cubert S5, Cubert X20), achieving processing times of < 170 ms and < 400 ms, respectively. We discuss camera characteristics, system parameters, and the suitability for clinical use and real-time applications.

## 1. INTRODUCTION

In open surgery settings, such as visceral surgeries, adequate tissue perfusion is critical for surgical success, especially for bowel anastomosis. Current assessments rely on the subjective judgment of the surgeon, leading to variability in surgical outcomes. Failure rates between 1% to 19% have been reported for bowel surgeries [1,2], leading to severe complications.

Hyperspectral imaging is a promising candidate for analyzing tissue perfusion non-invasively. It has been applied for numerous tasks in medical imaging [3]. Several methods have been demonstrated, including point-scanners, line-scanners, spectral scanning and snapshot systems [4]. The latter systems, such as mosaic cameras, are advantageous due to their real-time capability for medical imaging [5]. However, hyperspectral imaging in open surgery settings remains challenging. Currently, the surgical workflow must be interrupted to record hyperspectral images. The HSI system is brought in close proximity to the patient, the surgical lights and room lights are dimmed or switched off [6,7], HSI images are taken and then the surgical workflow is continued. The effort required for taking HSI images is substantial, and the imaging procedure disrupts the surgical workflow, leading to increased operating times.

To this end, we present a method for continuous integration of hyperspectral imaging into the surgical workflow. We demonstrate continuous perfusion assessment by using a hyperspectral light-field camera. Our system uses an efficient, correlation-based algorithm for processing light-field data with low computational effort for visualizing tissue oxygenation during surgeries. We show the capabilities of two distinct hyperspectral light-field cameras in our setup and subsequently evaluate both.

---

*stefan.kray@hs-pforzheim.de; https://www.hs-pforzheim.de/

## 2. METHODS AND MATERIALS

**Concept**

We propose a new approach for oxygen visualizations during surgery. The general concept of our approach involves a hyperspectral light field camera, which is positioned outside the sterile area of the operating room. For example, it can be mounted on the ceiling, integrated into the surgery light, or placed in the background on a tripod. It records images continuously during the extracorporal period of the surgery (Figure 1.)

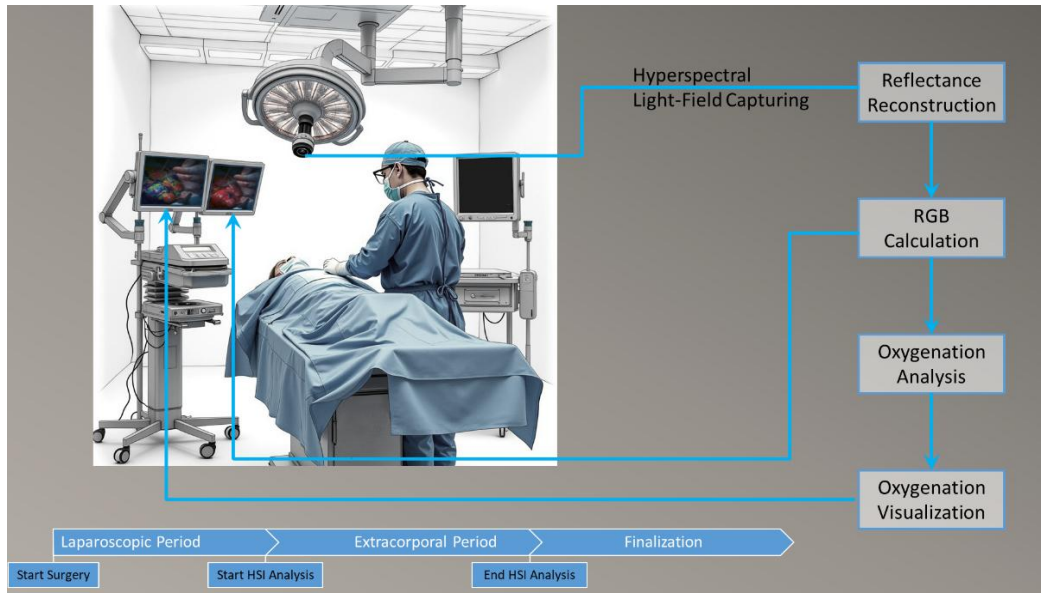

Figure 1. Concept and clinical setup for continuous perfusion assessment.

RGB images are calculated from the camera data. Values of tissue oxygenation are determined by the proposed correlation algorithm and the results are transformed into oxygenation images, which are continuously displayed without interrupting the surgery.

**Algorithm**

The first part of the algorithmic workflow replicates the original functionality of the manufacturer's software, serving as a foundation for a deep integration of tissue oxygenation analysis. A raw sensor image is transformed to a series of spectral images (raw data cube). Pre-calibrated homographies are applied to compute a transformed cube, ensuring alignment of each spectral image with the scene at the working distance. Dispersion maps as well as a scene dependent white reference cube are used to calculate the reflectance cube. Figure 2 shows this workflow.

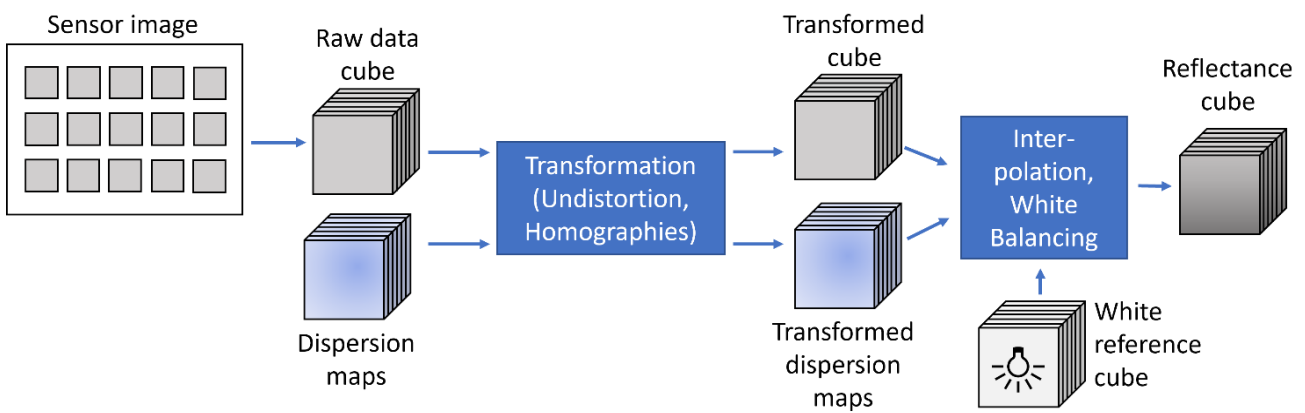

Figure 2. First step of the workflow for calculating reflectance cubes from the raw sensor image.

The pre-calibrated dispersion maps are also transformed to this working distance, assigning a fixed wavelength to each pixel. Linear interpolation is used to calculate a uniform hyperspectral cube. The hyperspectral data is normalized by a scene dependent white reference cube to produce the reflectance cube of the scene.

The proposed correlation algorithm to compute color-coded oxygenation images is presented in Figure 3. A 2D RGB image is calculated by evaluating the red, green, and blue color bands of the hyperspectral reflectance cube. For each pixel of the scene, a correlation between the spectral data and literature values for blood oxygenation is computed. Next, a color overlay representing tissue oxygenation values is determined.

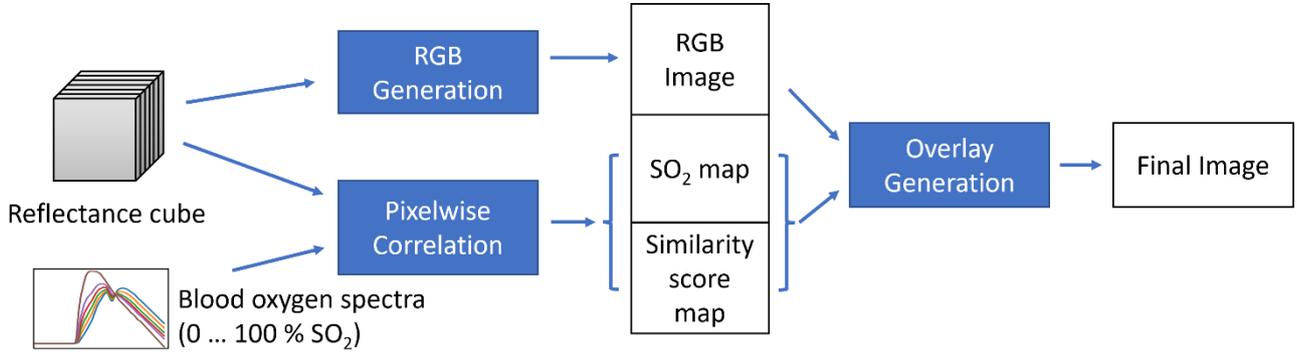

Figure 3. Second step of the workflow for generating colored oxygenation images from the reflectance cube.

Measured reflectance values $R_{cam}$ are compared to the literature $R_{SO_2}$ values across the wavelengths $\lambda_n$, using the spectral angle mapper (SAM) [8].

$$SAM_{SO_2}(x,y) = cos^{-1}\left(\frac{\sum_n R_{cam}(\lambda_n,x,y) \cdot R_{SO_2}(\lambda_n)}{\sqrt{\sum_n R_{cam}^2(\lambda_n,x,y)} \cdot \sqrt{\sum_n R_{SO_2}^2(\lambda_n)}}\right)$$

Assuming that ovine blood has sufficiently similar optical characteristics to human blood[9], we calculate a SAM score for each pixel $(x,y)$ and different $SO_2$ oxygenation levels. We assign the corresponding $SO_2$ value to each pixel position by choosing the blood reference curve $R_{SO_2}$, exhibiting the highest spectral similarity within these oxygenation levels. In this manner, we generate two values for each pixel. The first value is the best correlation match, representing the oxygenation level for the pixel. The second value is the correlation intensity, e.g., the SAM score itself, which helps us to differentiate between human tissue and background signal (e.g. surgical table, gloves, and other non-tissue elements). The $SO_2$ map is generated with the help of the best correlation match, the similarity score map is generated with the SAM score for each pixel. These two maps are then used to calculate a color overlay for the RGB image. Each $SO_2$ value is encoded using a color map (see Figure 4, right) for pixels identified as human tissue.

**Cameras**

For evaluation purposes, two hyperspectral light-field cameras are used simultaneously within this study. The first camera, Cubert Ultris S5[1], is a hyperspectral camera based on 42 microlenses in combination with a linear color filter. The camera, hereafter referred to as S5, generates 51 spectral bands ranging from 450 nm to 850 nm with an average spectral resolution of 8 nm. It has a spatial resolution of 290 x 275 pixels and is based on a 5-megapixel sensor. The camera has a quantization depth of 12 bits. It can be read out with a maximum frequency of 15 frames per second (fps). The field of view is 30 degrees.

The second camera, hereafter referred to as X20, is the Cubert Ultris X20 hyperspectral camera[1], based on 66 microlenses with individual color filters. It captures a light spectrum ranging from 350 nm to 1002 nm with 164 bands and an average spectral resolution of 4 nm. The spatial resolution is 410 x 410 pixels, using a 20-megapixel sensor. The camera also records with a quantization depth of 12 bits and supports a maximum frame rate of 8 fps. The field of view is 35 degrees.

---

[1] Cubert GmbH, Ulm, Germany

**Clinical Setup**

Our system is used intraoperatively during a bowel surgery. Each camera is connected to its own notebook and mounted on a tripod. They are positioned in the non-sterile area of the operating room and directed at the surgery field. The distance from the cameras to the surgical site, named as working distance in the following, is set to 56 cm.

The surgical light, having a spectrum ranging from 450 nm to 700 nm, is used as the light source for generating hyperspectral images. Due to the high illuminance >100.000 lux of the surgical light, images were recorded with low integration times of ≤5 ms for the S5 and ≤2 ms for the X20. The frame rate was set to 1 fps. A white sterile gauze was placed in the scene to capture the reflectance of the light source in the surgery field. This data served as a white reference cube for calculating reflectance cubes.

Our pilot study was approved by the Ethics Committee of the State Medical Association of Hesse (2024-3609-evBO) and conducted on a patient with Crohn´s disease. The patient underwent a robot-assisted ileocecal resection (da Vinci Xi, Intuitive Surgical, Sunnyvale, USA). Continuity was restored by a hand-sewn ileoascendostomy. Images were recorded during the extracorporal resection and anastomosis phases of the surgery.

## 3. RESULTS

Clinical results for visualizing tissue oxygenation by our method are shown in Figure 4. Figure 4(a) depicts the bowel segment shortly after it was extracted from the abdominal cavity. The tissue appears mostly red, representing high oxygenation levels, with a few parts appearing in blue. Figures 4(b) and 4(c) show the progression of the surgical resection. Further supplying arteries are interrupted and the bowel segment appears more blueish, as the blood supply is progressively cut off. Figure 4(d) represents the final image, shortly before the bowel segment is transected by the surgeon. The orange arrows indicate the resection line selected by the surgeon. The arrows correspond with the transition of red color to blue color within the image.

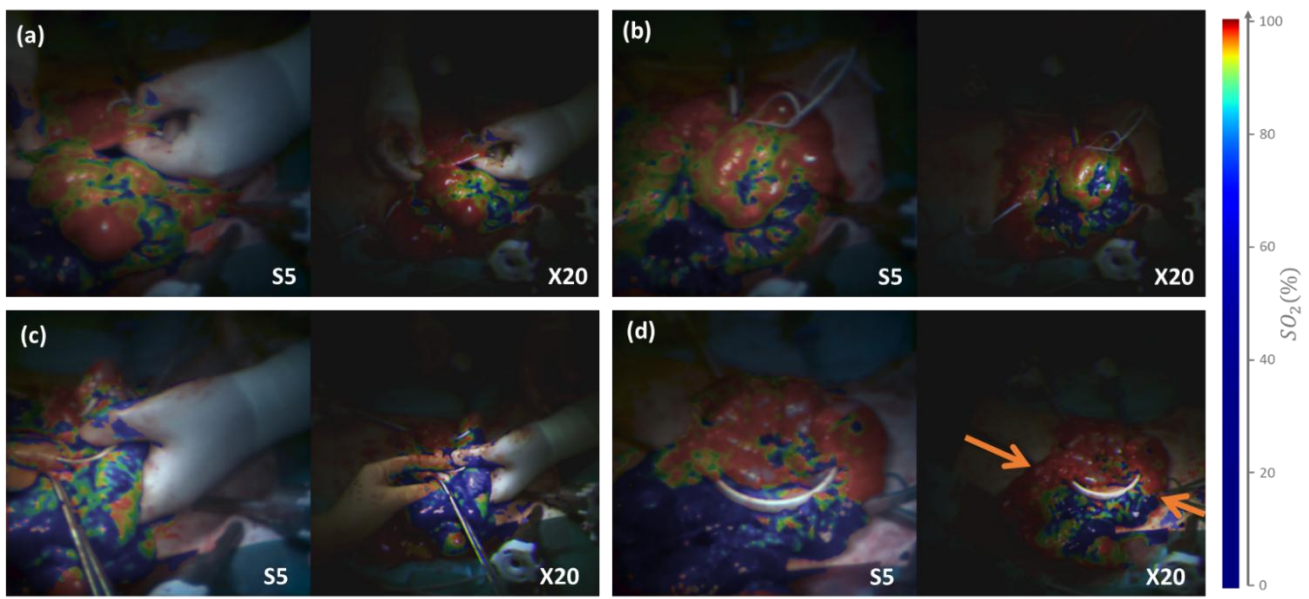

Figure 4. Oxygenation images of a bowel resection during surgery. Image (a) represents a scene at the beginning of the surgery, (b) and (c) show the progression of the bowel resection, (d) represents a scene shortly before the creation of ischemia for bowel anastomosis. The orange arrows indicate the resection line selected by the surgeon.

The algorithmic pipeline described in chapter 2 was implemented in our custom developed software, which continuously captures images during the surgery and supports the live display of oxygenation overlays. The resulting processing times, evaluated on a Thinkpad P14s notebook computer with an Intel Core Ultra 7 155H CPU, are shown in Table 1. Correlation timings were determined by comparing each pixel to 36 disctinct $SO_2$ oxygenation levels, providing an optimal resolution for high-quality colorizations. The S5 camera, equipped with a 5-megapixel sensor, achieves a processing time of 162 ms on CPU, most of which is spent on the calculation of the reflectance cube and the oxygenation correlation. The X20 camera is equipped with a 20-megapixel sensor, resulting in processing times of 397 ms on CPU. This is sufficient to display live images at 1-2 fps during surgery.

Table 1. Processing times for the S5 and X20 camera, evaluated on a notebook computer (Thinkpad P14s) equipped with a Intel Core Ultra 7 155H CPU and 64 GB RAM. All values in milliseconds (ms).

|  | CPU S5 | CPU X20 |
| ---: | :---: | :---: |
| Reflectance Cube | 57,69 | 185,95 |
| RGB Image | 10,25 | 75,17 |
| Oxy Correlation | 80,98 | 104,16 |
| Oxy Image | 6,85 | 14,22 |
| Add. Overhead | 5,76 | 17,68 |
| Total (ms) | 161,53 | 397,18 |

## 4. DISCUSSION

Within this study, two different hyperspectral light-field cameras were used to analyze tissue oxygenation during bowel surgeries. The S5 camera has lower spatial and spectral resolution compared to the X20 camera but offers faster readout and has shorter processing time. The choice of the appropriate camera depends on the application area and resulting parameters, such as spectral resolution, spatial resolution, measurement depth, and required frame rate.

Both cameras are suitable for measuring tissue oxygenation, as demonstrated in Figure 4. and both cameras yield similar results. The proposed correlation method can be efficiently implemented, resulting in consistent outcomes. The resection line chosen by the surgeon aligns well with the oxygenation boundary, indicated by the red and blue regions of the bowel segment. Although validation with absolute oxygenation values was not conducted within this study, it is evident that our method is suitable for differentiating between perfused and non-perfused segments, which is crucial for creating ischemia and performing subsequent anastomosis.

Calculating reflectance cubes relies on accurate white reference calibration. We establish stable conditions by placing a white gauze into the surgery field and use this as a region of interest to calculate the spectrum of the surgical light. In this manner, we can measure the reference spectrum under changing lighting conditions. However, more complex calibration schemes may be beneficial for more challenging conditions [10].

Further effort is necessary to exploit the full potential of the high-resolution image sensors. By adapting an upsampling scheme [11,12], increased resolution for both cameras could be achieved in the future. Additionally, 3D information can be leveraged by using different perspectives to the scene due to the light-field technology. Implementing GPU acceleration will help to keep processing times low.

## 5. CONCLUSION

In this study, we demonstrated continuous perfusion assessment during bowel surgeries. Using hyperspectral light-field cameras in combination with efficient visualization algorithms, we were able to identify the resection line as chosen independently by the surgeon. Our results indicate that hyperspectral light-field imaging has the potential to be an objective assistance for assessing blood perfusion in surgical settings.

## ACKNOWLEDGEMENTS

This research was created as part of the 'Neospek' project. The project was made possible by funding from the Carl-Zeiss-Stiftung, No. P2022-07-006.